\documentclass[aip,jcp,reprint]{revtex4-1}

\usepackage{graphicx}
\usepackage{amsmath}

\newcommand{\COtwo}{CO$_2$}
\newcommand{\HtwoO}{H$_2$O}
\newcommand{\Htwo}{H$_2$}

\begin{document}

\title{NMR study of small molecule adsorption in MOF-74-Mg}

\author{M.\,G. Lopez}
\affiliation{Department of Physics, Wake Forest University,
Winston-Salem, NC 27109, USA.}

\author{Pieremanuele Canepa}
\affiliation{Department of Physics, Wake Forest University,
Winston-Salem, NC 27109, USA.}

\author{T. Thonhauser}
\email[E-mail: ]{thonhauser@wfu.edu}
\affiliation{Department of Physics, Wake Forest University,
Winston-Salem, NC 27109, USA.}

\date{\today}

\begin{abstract}
We calculate the carbon nuclear magnetic resonance (NMR) shielding for
CO$_2$ and the hydrogen shieldings for both H$_2$ and H$_2$O inside the
metal organic framework MOF-74-Mg. Our \emph{ab initio} calculations are
at the density functional theory level using the van der Waals including
density functional vdW-DF. The shieldings are obtained while placing the
small molecules throughout the structure, including the calculated
adsorption site for various loading scenarios. We then explore
relationships between loading, rotational and positional
characteristics, and the NMR shieldings for each adsorbate. Our NMR
calculations show a change in the shielding depending on adsorbate,
position, and loading in a range that is experimentally observable. We
further provide a simple model for the energy and the NMR shieldings
throughout the cavity of the MOF. By providing this mapping of shielding
to position and loading for these adsorbates, we argue that NMR probes
could be used to provide additional information about the position at
which these small molecules bind within the MOF, as well as the loading
of the adsorbed molecule.
\end{abstract}

\maketitle

\section{Introduction}\label{sec:introduction}

Metal Organic Frameworks (MOFs)\cite{Chen01, Eddaoudi00, Rosi03} have
become very popular over the last decade, as is evident by their
prevalence in recent studies and generous review in the
literature.\cite{James03, Ferey08, OKeefe09, Perry09, Han09, Morris10,
Furukawa10, Bordiga10} This interest is largely due to the wide range of
applications that have been identified for MOFs, ranging from molecular
gas storage (CH$_4$,\cite{Eddaoudi02, Ma08, Wu09} N$_2$,\cite{Nelson09,
Sumida10} \COtwo,\cite{Sumida10, Dietzel08, DAlessandro10, Millward05}
H$_2$\cite{Dinca06, Forster06, Vitillo08, Murray09, Hu10}) to gas
separation,\cite{Li09, Britt09, Sato10, Shimomura10, Ferey11} drug
delivery,\cite{Xiao07, Horcajada10} sensing,\cite{Xie10, Qiu09}
catalysis,\cite{Forster03, Kesanli03, Lee09, Ma09, Corma10, Liu10rev,
Liu10} and photo\-ca\-taly\-sis.\cite{Splan04, Silva10} The utility of
MOFs comes from their interactions with small molecules such as \Htwo,
\COtwo, and \HtwoO.  It is thus critical to understand the details of
the binding process when a small molecule is adsorbed into the MOF. To
this end, IR and Raman spectroscopy have been used extensively to study
small molecule adsorption in MOFs,\cite{Tan12, Nijem12, Nijem12-2} but
it can be difficult to determine where the reactive sites reside under
different loading scenarios.  Also, for these probes the strong signals
originating from the vibrational modes of the gas present in the
experiment chamber and the MOF itself can often dominate the spectrum,
making the analysis of the weak adsorbate signal challenging.  In the
following, we argue that NMR---which has already been used successfully
to study MOFs in a number of cases\cite{Stallmach99, Geier04,
Gonzalez05, Stallmach06, Gassensmith11, Kong12}---can be used to
facilitate a more detailed understanding of the static behavior of
MOF/adsorbate interactions and binding under various conditions. In
particular, we show that NMR can provide information about the position
at which these small molecules bind within the MOF, as well as the
loading of the adsorbed molecule.

In this work, we consider the particular MOF structure
MOF-74-Mg,\cite{Dietzel08-2} which has been shown to have very high
efficiency when capturing \COtwo,\cite{Britt09} a key property for gas
separation and storage applications.  For the small molecule adsorbed in
the MOF we consider \Htwo, \COtwo, and \HtwoO. The first is obviously
interesting for hydrogen-storage applications, while the second one is
of interest in carbon-capture applications. However, water by itself is
not necessarily interesting for applications, were it not for the fact
that it strongly impedes the performance in the first two cases. In
other words, the presence of water, due to its strong binding
characteristics, decreases the performance of MOFs in hydrogen-storage
and \COtwo\ capture applications, such that its careful study is
warranted.\cite{Canepa12}

The three molecules investigated in this study, i.\,e.\ \Htwo, \COtwo,
and \HtwoO, bind inside the MOF through phy\-sisorption. Thus, it is
apparent that the proper inclusion of van der Waals interactions is
crucial for the entire study. Therefore, we use density functional
theory (DFT), utilizing the van der Waals including functional
vdW-DF\cite{Dion04, Thonhauser07, Langreth09} to map the shielding of an
adsorbed molecule  within MOF-74-Mg to various characteristics.  This
truly non-local exchange-correlation functional has already successfully
been applied to study small molecule adsorption in a variety of
MOFs.\cite{Canepa12, Sagara05, Walker10, Nour11, Kong11, Centrone05,
Kuc08, Tan12, Nijem12, Nijem12-2, Nijem10, Yao12}

\section{Computational Details}\label{sec:comp_details}

The interaction of \Htwo, \COtwo, and \HtwoO\ with MOF-74-Mg was studied
using DFT with vdW-DF as implemented in
\textsc{QuantumEspresso}.\cite{Giannozzi09} It is well known that
binding distances will usually be slightly overestimated using
vdW-DF.\cite{Langreth09} We used the primitive rhombohedral unit cell of
MOF-74-Mg with space group R$\bar{3}$ and 54 atoms. The initial geometry
of MOF-74-Mg was relaxed, fixing the lattice parameters according to the
experimental values of $a = 15.117$~\AA\ and $\alpha =
117.742^{\circ}$.\cite{Wu09} A complete volume relaxation for all
loadings considered in this paper would have been extremely
computationally expensive, so that we fixed the lattice constants to the
experimentally measured ones.  In our testing, for the expected worse
case of 12 \HtwoO\ molecules adsorbed in MOF-74-Mg, we find that there
is only a 1\% change in the channel diameter when a full relaxation is
performed. For each adsorption case we have relaxed the internal
coordinates until the total force was below
1$\times$10$^{-4}$~Ry~Bohr$^{-1}$.  Ultrasoft pseudopotentials together
with a plane-wave cutoff of 35 Ry were used to describe the wave
functions, while the charge-density cutoff was set to 280~Ry.  The
convergence threshold for the self-consistency of the total energy was
set to 5$\times$10$^{-11}$ Ry, ensuring an accurate sampling of the
complex potential energy surface for MOF-74-Mg.  

With the coordinates obtained from the geometry relaxation, the
adsorption energies and NMR shielding parameters were calculated using
norm-conserving GIPAW pseudopotentials,\cite{Pickard01} which allow for
the wavefunction reconstruction in the atomic core region. Structural
aspects are not so sensitive to the cutoffs, ensuring the lower values
reported above for the geometry optimizations are appropriate. But, the
NMR shielding parameters are much more sensitive, so that higher values
are needed. Accordingly, we used a plane-wave cutoff of 120~Ry and a
charge-density cutoff of 420~Ry, resulting in a convergence of the
absolute shielding to within 0.05~ppm.  However, in this study we are
mostly interested in the \emph{change} in shielding of the adsorbed
molecule compared to its gas phase, which is converged to within less
than 0.001~ppm.  For the NMR shielding calculations,\cite{foot1} we used
a combination of the linear-response\cite{Mauri96} and new
converse\cite{Thonhauser08, Thonhauser09, Ceresoli10} methods---the
latter being built entirely on the theory of orbital
magnetization.\cite{Thonhauser11, Thonhauser05, Ceresoli06, Resta_2005}
The adsorption energies reported in this study were calculated with
identical parameters to these NMR calculations.

In addition, we cross-checked our calculated NMR shieldings for selected
adsorption cases of H$_2$ with \textsc{Vasp}\cite{Kresse94} 
(a plane-wave code) and \textsc{Gaussian}\cite{Frish09}
localized basis-set code). We find that the gas-phase shieldings of
those codes agree to within less than 0.1~ppm with our results, but more
importantly, the shielding difference between gas-phase and adsorbed
molecules agrees with our calculations to within 0.01 ppm.

\section{Results}\label{sec:results}

\subsection{Reactivity and binding energy}\label{sec:binding}

\begin{figure*}
\includegraphics[width=\textwidth]{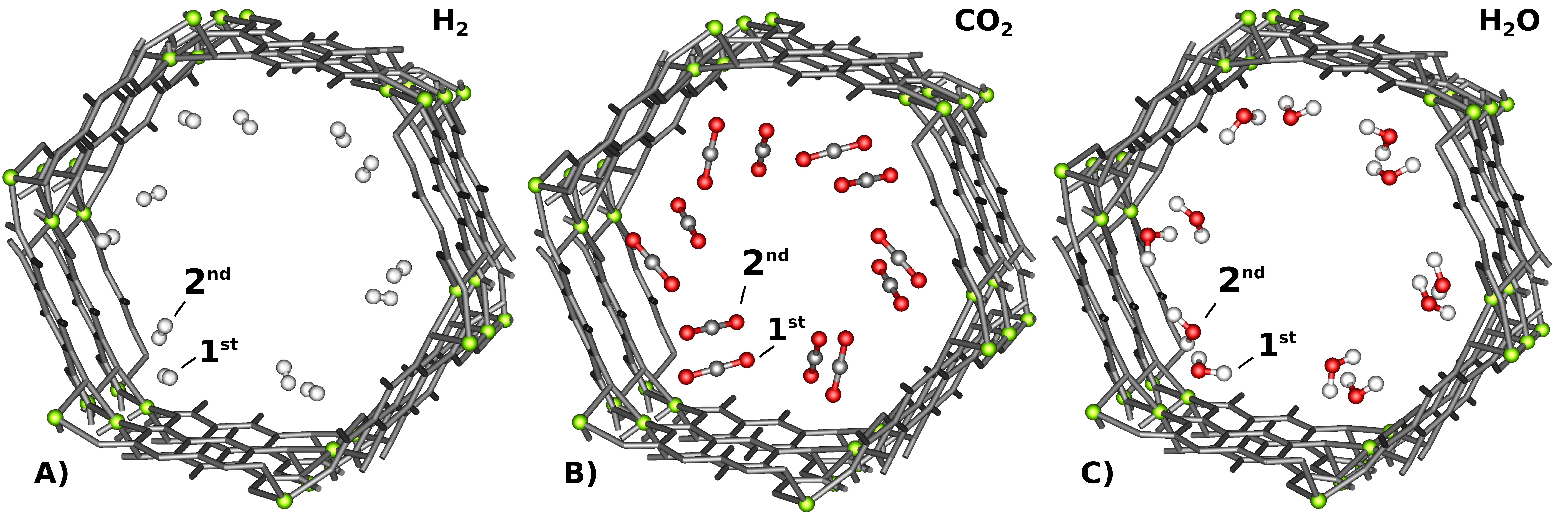}
\vspace{-5ex}
\caption{\label{fig:loadings} 
The MOF-74-Mg structure is shown as sticks with metal Mg ions
highlighted as green balls.  The three panels show
\textbf{A)} \Htwo,
\textbf{B)} \COtwo, and 
\textbf{C)} \HtwoO\
as ball-and-stick representations in the case of very high loading with
all six primary and all six secondary binding sites occupied, labelled
as $1^\text{st}$ and $2^\text{nd}$, respectively.  Notice the classic
dimer configuration of the adsorbed water molecules.}
\end{figure*}

Although the subject of the binding characteristics itself is not the
main focus of this study, we reproduce and extend results here that have
been published elsewhere\cite{Canepa12} but are nonetheless important
for our NMR study. In particular, the adsorption energy of \Htwo,
\COtwo, and \HtwoO\ in MOF-74-Mg under different loading situations is
of interest, as it defines the primary and secondary binding sites and
binding geometries, for which we will report NMR results below. The
structure of MOF-74, which can be seen in Figs.~\ref{fig:loadings} and
\ref{fig:co2_rot}, consists of hexagonal channels, where metal atoms at
the corners are connected by benzenedicarboxylate linkers.  The primary
binding sites are located near the six metal ions in each unit cell,
while secondary binding sites are nearer to the linkers. For further
details, see Ref.~\onlinecite{Canepa12}.

The adsorption energy, $\Delta E$, of a guest molecule M in the MOF is
defined as
\begin{equation}
\label{eq:be}
\Delta E = E_\text{MOF+M} - [ E_\text{MOF} + E_\text{M}]\;,
\end{equation} 
where  $E_\text{MOF}$ and  $E_\text{M}$ are the energies of the MOF and
the molecule in their fully relaxed form, and $E_\text{MOF+M}$ is the
energy of the MOF with the adsorbed M. Results for the adsorption
energies $\Delta E$ are given in Table~\ref{table:be} for several
different loadings: (i) \emph{low loading}, i.e.\ one guest molecule per
cell occupying a primary binding site; (ii) \emph{high loading}, six
guest molecules per cell completely saturating all available primary
sites; (iii) \emph{high loading}, 7 guest molecules per cell completely
saturating all primary sites and one secondary site; and (iv) \emph{very
high loading}, 12 molecules per unit cell, occupying all available
primary and secondary binding sites. For a depiction of the binding
geometries in those cases, see Fig. \ref{fig:loadings}. We find good
agreement with the experimental adsorption energies of --0.11 $\pm$
0.003 eV for H$_2$\cite{Zhou08} and --0.49 $\pm$ 0.010 eV for
CO$_2$,\cite{Valenzano10} attesting to the importance of correctly
including van der Waals interactions in these simulations. In a recent
study\cite{Canepa12} we also computed vibrational frequencies to obtain
the thermal and zero-point energy (ZPE) corrections to these adsorption
energies, allowing for a more accurate comparison to measured adsorption
heats. But, for the cases considered here, we found that these
corrections are on the order of 0.01~eV or less, so they are not again
reported here.

\begin{table}
\caption{\label{table:be}Adsorption energies $\Delta E$ of molecules in
MOF-74-Mg in eV for different loadings.  $\Delta E_{\rm prim}$ and
$\Delta E_{\rm sec}$ are the average adsorption energies for the
molecule at the primary and secondary binding sites, respectively.
$\Delta E_{\rm avg}$ is the average adsorption energy per molecule
considering all adsorbed molecules together.}
\begin{tabular*}{\columnwidth}{@{\extracolsep{\fill}}lcccr@{}}\hline\hline
M       & Loading & $\Delta E_{\rm prim}$ &$\Delta E_{\rm sec}$ & $\Delta E_{\rm avg}$ \\\hline
H$_2$  & 1  & --0.15  & n/a    & --0.15 \\
       & 6  & --0.16  & n/a    & --0.16 \\
       & 7  & --0.16  & --0.12 & --0.15 \\
       & 12 & --0.16  & --0.12 & --0.14 \\
CO$_2$ & 1  & --0.50  & n/a    & --0.50 \\
       & 6  & --0.50  & n/a    & --0.50 \\
       & 7  & --0.50  & --0.43 & --0.49 \\
       & 12 & --0.50  & --0.46 & --0.48 \\
H$_2$O & 1  & --0.79  & n/a    & --0.79 \\
       & 6  & --0.76  & n/a    & --0.76 \\
       & 7  & --0.74  & --0.61 & --0.72 \\
       & 12 & --0.74  & --0.65 & --0.70 \\\hline\hline
\end{tabular*}
\end{table}

The influence of ``crowding'' on the adsorbed molecules in high-loading
situations is present, but not dominating. For \Htwo, the contribution
of the lateral interactions in the high-loading scenarios is negligible,
being less than $7\%$ of the total binding energy in the case of 12
adsorbed \Htwo\ molecules. For the \COtwo\ and \HtwoO\ molecules, the
lateral interactions (attractive) contribute less than $10\%$ of the
total binding energy as given in Table~\ref{table:be} when only each of
the six primary binding sites is occupied. However, when all available
primary and secondary sites are occupied (12 molecules bound), this
contribution increases to $18\%$ and $25\%$ for \COtwo\ and the hydrogen
bonding \HtwoO\ cases, respectively.

\subsection{NMR -- Loading study}

We now move to the main topic of this study---the analysis of the NMR
chemical shielding of \Htwo, \COtwo, and \HtwoO\ in MOF-74-Mg. Unless
otherwise stated, the reported values in parts per million (ppm) are the
\emph{change} in isotropic NMR chemical shielding $\Delta \sigma$ when
the gas-phase molecule M is adsorbed in the MOF, i.\,e.
\begin{equation}
\label{eq:del_shield}
\Delta \sigma =  \sigma_\text{M in MOF}-\sigma_\text{M in gas phase}\;.
\end{equation}
After determining the primary and secondary binding sites for the three
adsorbate molecules in the different loading cases (see
Sec.~\ref{sec:binding}), we calculated the NMR shielding of the adsorbed
molecules at those positions within the MOF-74-Mg structure;
corresponding values are reported in Table~\ref{tab:nmr}.  In primary
and secondary site high-loading cases, the average of equivalent atoms
is reported.  When a single secondary site is occupied by water, it
forms a classic water dimer by hydrogen bonding with the water at the
nearest primary binding site. For this reason, only the five
unpaired primary sites are averaged in Table~\ref{tab:nmr} for \HtwoO\
with a loading of 7; the values of the ``special'' molecule, to which
the seventh molecule in the secondary binding site attaches, are
reported separately.

The chemical shielding dependence on adsorbate loading can be seen in
Table~\ref{tab:nmr} for all three molecules.  It is perhaps not
surprising to see that while \Htwo\ shows a typically weak vdW
physisorption-like interaction with the MOF (also indicated by the
smaller binding energies in Table~\ref{table:be}), water displays the
more typical proton NMR deshielding behavior. When \HtwoO\ occupies any
of the secondary binding sites, it assumes a dimer configuration with
the molecule in the primary site with an average hydrogen-bond distance
of 1.86~\AA{} which is consistent with previous results.\cite{Kolb11}
This hydrogen bonding produces a large effect which can be seen by large
shielding changes for \HtwoO\ in high-loading cases shown in
Table~\ref{tab:nmr}.  These large shielding changes are in good
agreement with previous NMR studies of liquid water in which hydrogen
bonding also plays a significant role.\cite{Thonhauser09,foot2} While it is
conceivable that as the loading increases, the water molecules could
form a hydrogen-bond chain instead of occupying the remaining secondary
binding sites in the cell, calculations and experiment\cite{Kolb11}
indicate that the hydrogen bond only accounts for about one third of the
secondary site binding energy, making this configuration less favorable.  

\begin{table}
\caption{\label{tab:nmr}Change in NMR shielding $\Delta \sigma$ in ppm
(carbon for \COtwo, hydrogen for \Htwo\ and \HtwoO) upon adsorption
relative to the gas phase for the primary and secondary binding sites.
Shieldings for the primary and secondary binding sites are given in
separate columns. For \Htwo\ and \HtwoO\ the values for both hydrogens
are given in separate lines.  In the cases of \Htwo\ and \HtwoO, the
first row is the hydrogen which is more strongly interacting with the
oxygen plane of the metal binding site complex, while the second is
further away.  For higher loading situations, the shieldings have been
averaged over all equivalent sites.} 
\begin{tabular*}{\columnwidth}{@{\extracolsep{\fill}}lccc@
{\hspace{-1.2em}}cc@{\hspace{-1.2em}}r@{}}\hline\hline
Loading & 1       & 6       & \multicolumn{2}{c}{7} & \multicolumn{2}{c}{12}\\
        & prim.   & prim.   & prim.  & sec.   & prim.  & sec.  \\\hline
\Htwo   & 0.48    & 0.50    & 0.54   & 0.09   & 0.25   & 0.14  \\
        & 0.24    & 0.30    & 0.26   & 0.50   & 0.51   & 0.58  \\
\COtwo  & 1.01    & 2.42    & 2.16   & 0.95   & 2.69   & 0.81  \\
\HtwoO  & --0.83  & --0.85  & --1.06\footnotemark[1]\footnotetext[1]
        {--0.57 and --4.99 ppm for the molecule to which the seventh \HtwoO\ attaches.}
        & --4.41 & --0.53 & --3.65\\
        & --0.74  & --0.56  & --0.78\footnotemark[1]
        & --0.37 & --5.42 & --0.12\\\hline\hline
\end{tabular*}
\end{table}

From these results, it can be seen that---while direct usage of NMR alone
to determine relative loadings for \Htwo\ might be difficult---the
situation is more positive for \COtwo\ and \HtwoO.  For \COtwo, relative
loadings of the primary binding site show up clearly as a significant
difference.  While the differences are not as obvious for the secondary
site, the presence of two peaks---one around 1~ppm and another greater
than 2~ppm from the gas phase shieldings is a good indication of \COtwo\
occupying both primary and secondary sites.  For \HtwoO, the changes in
shieldings at the primary binding site (in the absence of hydrogen
bonding) are much smaller as loading increases, but the presence of low
loadings can still be detected with a change in shielding around 0.8~ppm
less than the gas phase value.  Furthermore, the formation of hydrogen
bonds when greater than six molecules are adsorbed is a clear indication
of high-loading situations for \HtwoO.

\subsection{NMR -- Rotational study}\label{sec:rot}

\begin{figure}
\begin{center}
\includegraphics[width=0.7\columnwidth]{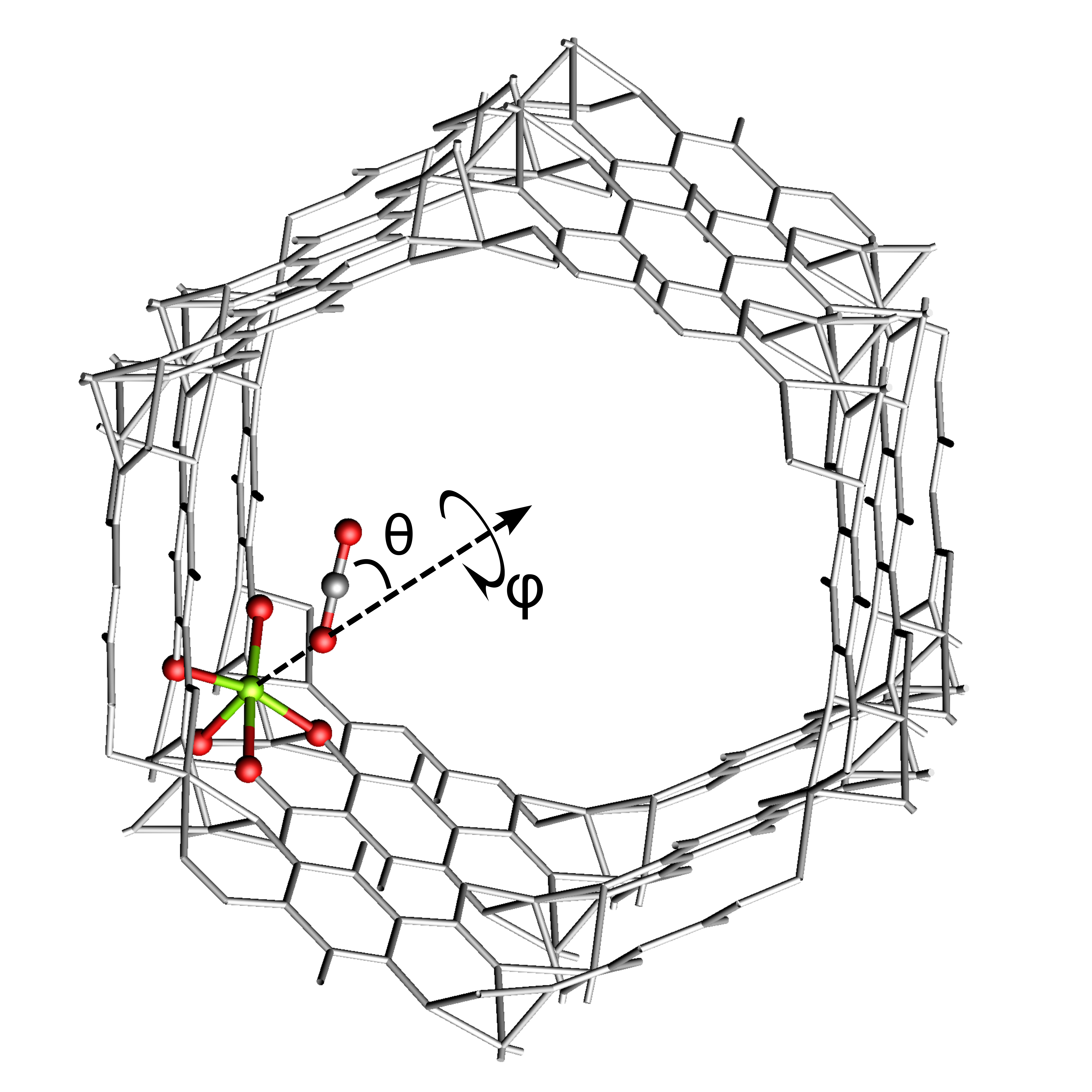}
\vspace{-4ex}
\end{center}
\caption{\label{fig:co2_rot} Schematic drawing defining the rotation and
its axis, depicted for the case of \COtwo.  The molecule is placed at
one of six identical primary binding sites, emphasized by the colored
ball-and-stick representation with carbon, oxygen and magnesium
represented as grey, red, and green balls, respectively.  The axis of
rotation is the polar axis defined as a line from the binding Mg through
the closest atom of the small molecule. The molecule is then rigidly
rotated (indicated by the curved arrow) about this axis, sampling the
azimuthal angle $\varphi = 0-360^\circ$ and keeping $\theta$ fixed at its
computed value in the lowest energy configuration---approximately
$94^\circ$ and $55^\circ$ for \Htwo\ and \COtwo, respectively.} 
\end{figure}

\begin{figure}
\includegraphics[width=\columnwidth]{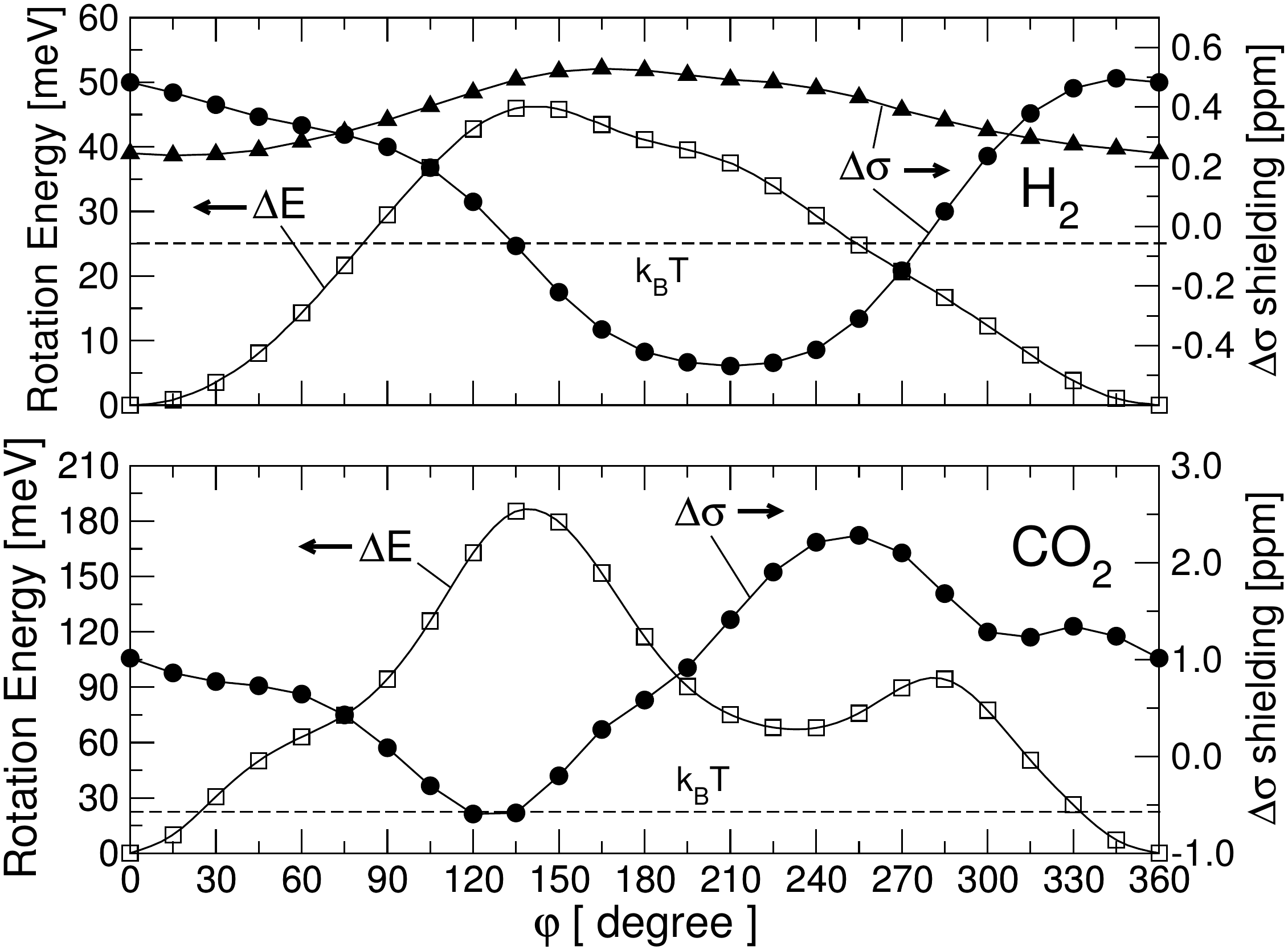}
\caption{\label{fig:co2_rot_plot} The change in energy and NMR shielding
calculated for \Htwo\ and \COtwo\ as a function of rotation at the
primary binding site in MOF-74-Mg. For \Htwo, the shielding for both
atoms is shown, with the fixed hydrogen shown as solid triangles and the
rotated hydrogen with solid circles.  For a definition of the angle and
axis of rotation, see Fig.~\ref{fig:co2_rot}. The energy scale is shown
on the left and the NMR shielding scale is on the right. The dashed
lines indicate $k_BT$ at room temperature, i.\,e.\ 25.6~meV.}
\end{figure}

Strictly speaking, Table \ref{tab:nmr} gives the NMR shielding change
for molecules at the primary and secondary binding sites at zero
temperature.  However, for finite temperatures the molecules will start
to ``wiggle''---governed by the potential energy surface around the
binding site---resulting in small changes in shielding.  In order to
investigate the order of magnitude of such changes, we studied the
rotational (and in the next section, translational) behavior of \Htwo\
and \COtwo\ at the primary binding site.  The axis of rotation is
defined by an imaginary line between the Mg atom and the closest
hydrogen in \Htwo\ or the closest oxygen in \COtwo\ at the binding site,
as shown in Fig.~\ref{fig:co2_rot} for \COtwo.  The binding geometry is
the lowest energy geometry and defines the zero-degree configuration for
the azimuthal angle $\varphi$ in Figs.~\ref{fig:co2_rot} and
\ref{fig:co2_rot_plot}. The relative angle $\theta$ was then ``frozen''
and the molecule was rotated from $\varphi=0^\circ$ to $\varphi=360^\circ$ in
intervals of $15^\circ$. Note that this rotation is not meant to
accurately sample energies and shieldings of physically likely
situations---rather, it should give an estimate for the sensitivity of
these properties in close proximity of the binding site.

In Fig.~\ref{fig:co2_rot_plot} we show both the energy and NMR shielding
as the molecules are rotated $360^\circ$.  Our results show that for
\Htwo\ at room temperature, there is a variation in the shielding by as
much as 0.8~ppm, and at elevated temperatures as much as 1~ppm.  For
\COtwo\ at room temperature, there is a variation in the shielding up to
0.4~ppm, but at very high temperatures this variation can grow as large
as 2.5~ppm.  Note that there is a secondary minimum in the rotational
energy for \COtwo\ around 230$^\circ$, so that low temperature
measurements could detect a secondary peak around 2~ppm further from the
gas-phase shift. This secondary minimum has a depth of 30 meV and thus
can maintain trapped molecules in this rotational configuration at room
temperature.

\subsection{NMR -- Positional study}

\begin{figure}
\begin{center}
\includegraphics[width=0.8\columnwidth]{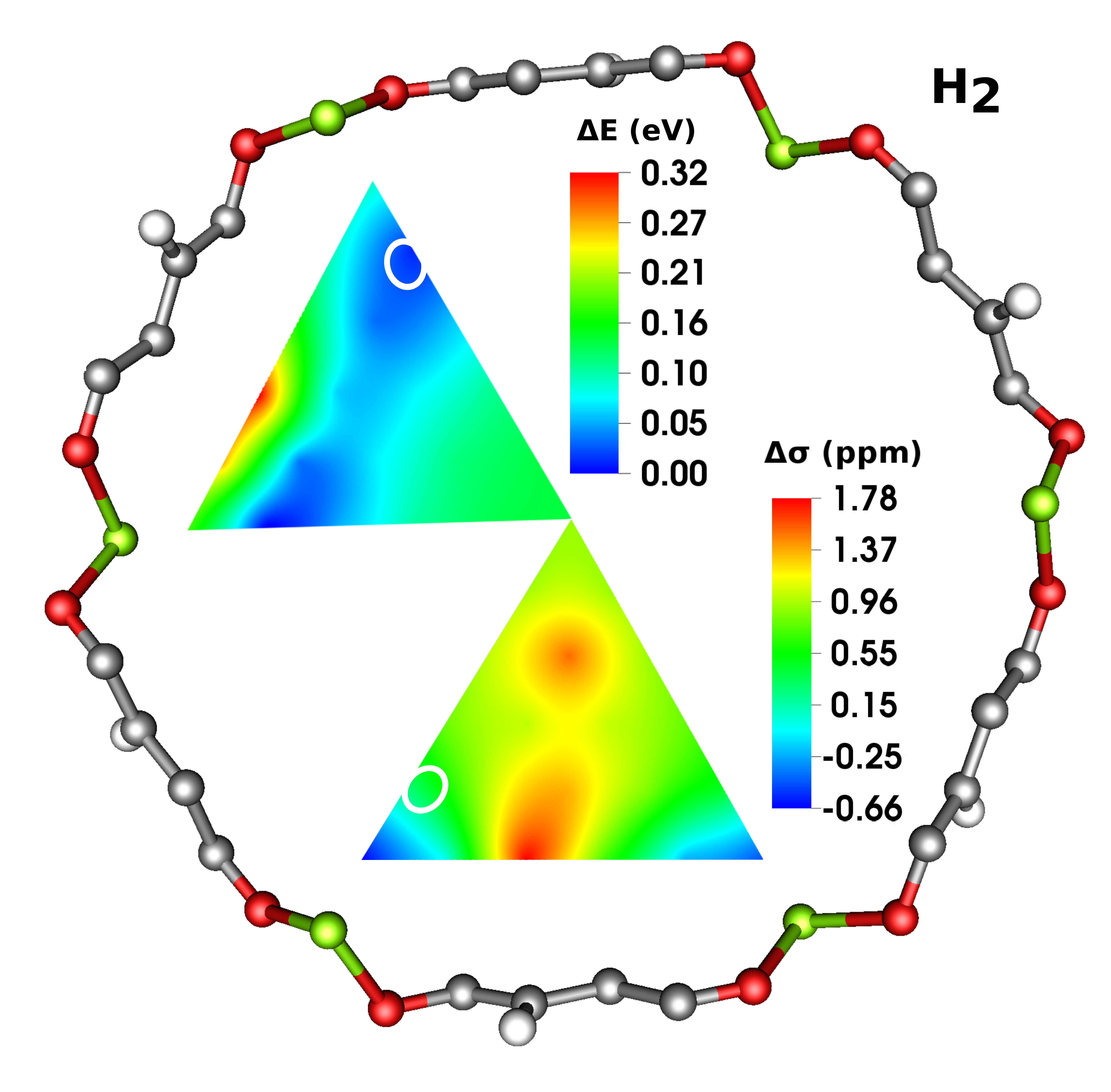}
\vspace{-5ex}
\end{center}
\caption{\label{fig:h2} Each triangle shows a map of the change in i)
energy and ii) the NMR shielding as the \Htwo\ molecule is moved
throughout the cavity of MOF-74-Mg. The MOF is oriented so that the
viewer is looking along the direction of the channel, and hydrogen,
carbon, oxygen, and magnesium atoms are represented as white, grey, red,
and green balls, respectively. The two maps were calculated at identical
points, but they are shown here in symmetry-equivalent locations within
the MOF structure for better comparison.  The shielding is averaged over
both hydrogen atoms. The energy is plotted relative to the binding
energy. The white circle indicates $k_BT$ at room temperature, i.\,e.\
25.6~meV, and is thus an estimate for the region around the primary
binding site accessible through translational fluctuations at room
temperature.}
\end{figure}

\begin{figure}
\begin{center}
\includegraphics[width=0.8\columnwidth]{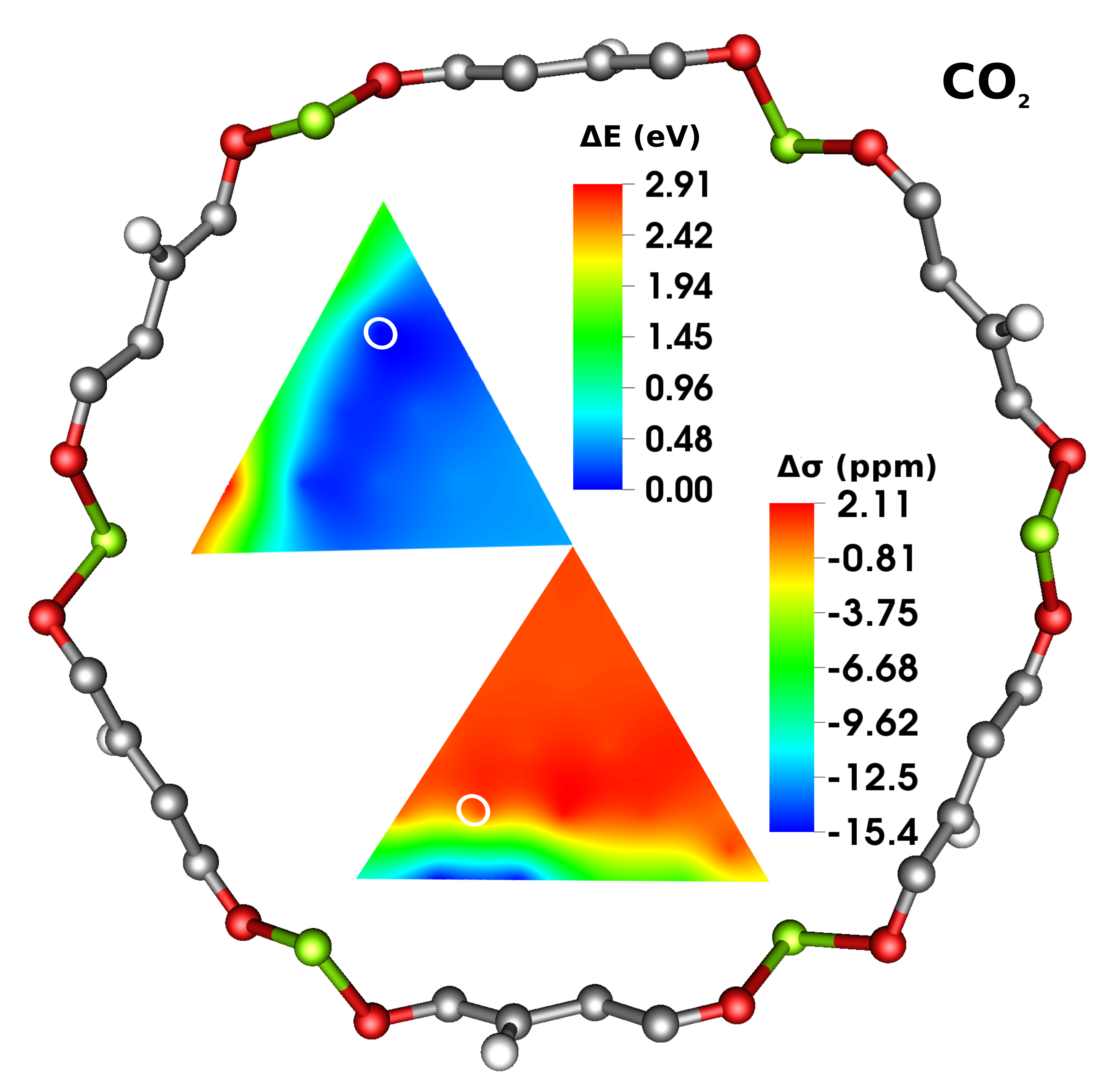}
\vspace{-5ex}
\end{center}
\caption{\label{fig:co2}
Same as Fig.~\ref{fig:h2}, but here for \COtwo.}
\end{figure}

\begin{figure}
\begin{center}
\includegraphics[width=0.8\columnwidth]{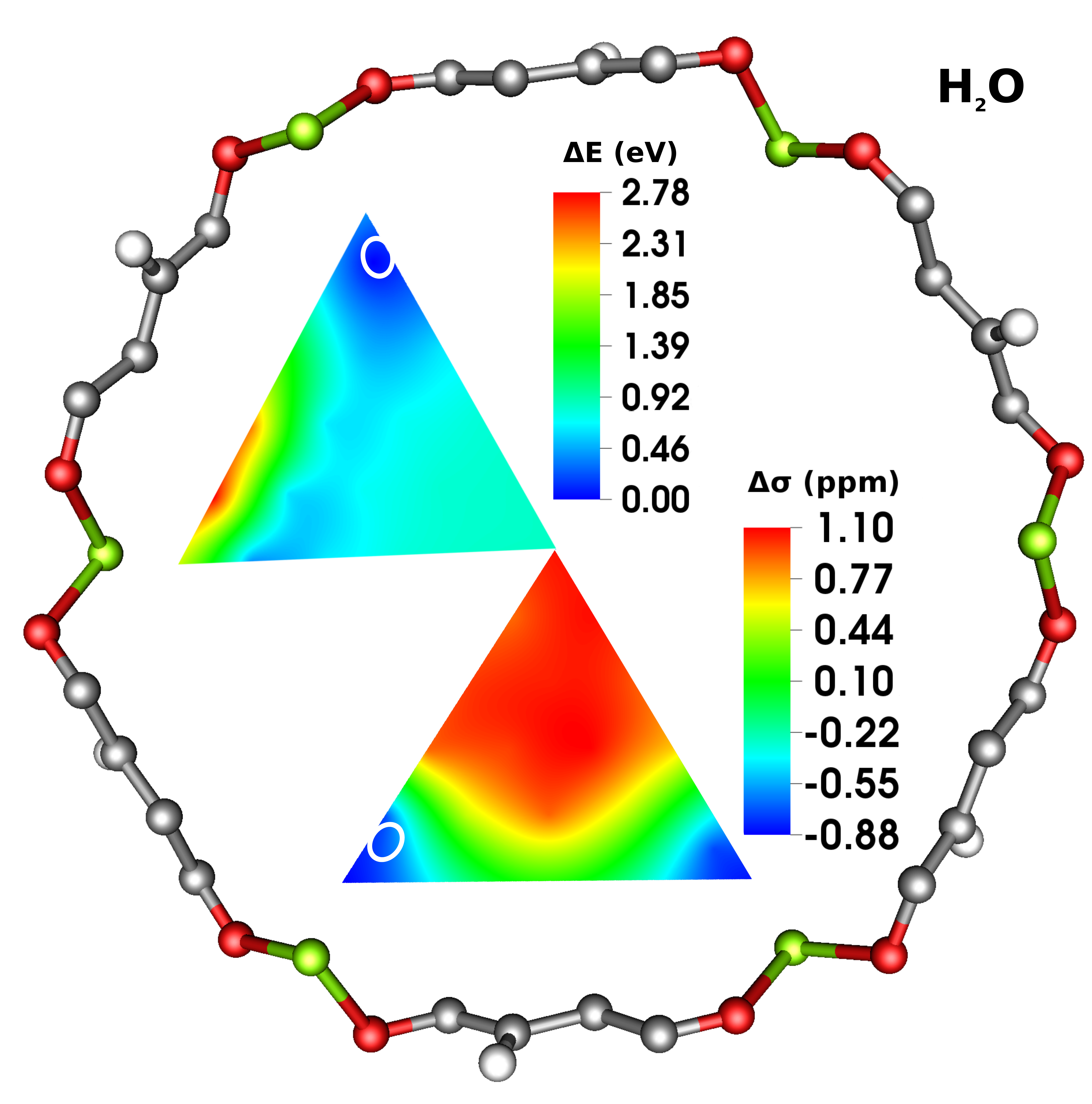}
\vspace{-5ex}
\end{center}
\caption{\label{fig:h2o}
Same as Fig.~\ref{fig:h2}, but here for \HtwoO.}
\end{figure}

In the previous section we investigated the sensitivity of energy and
shielding in the proximity of the binding site for simple rotations. In
the following, we investigate the same for large deviation from the
binding site---in fact, we calculate the energy and shielding change
throughout the entire cavity of the MOF.  In Figs.~\ref{fig:h2},
\ref{fig:co2}, and \ref{fig:h2o} we show a map of the energies and NMR
shieldings calculated for the small molecules at different positions
within the cavity of MOF-74-Mg. The values shown are the change in
shielding when the molecules are taken from gas phase into the MOF in a
low-loading scenario.  The planes over which the shielding was studied
are defined by three points: the center of mass of the MOF-74-Mg unit
cell, the binding site of the small molecule as calculated by the
geometry relaxations described in Sec.~\ref{sec:comp_details}, and the
Mg atom at which the molecule is adsorbed.  Note that a plane defined in
this way is not coplanar with the MOF structure, the normal vector to
the plane being slightly tilted (by $10^\circ-15^\circ$) away from the
channel direction.  This implies that replicating the NMR maps displayed
in these figures using the $D_{3d}$ symmetry of the MOF would yield a
slightly discontinuous image.  The values were calculated at 26
equidistant points within this plane and linear interpolation was used
to make the complete map.  At each of the sampled points, the center
atom of the molecule was ``pinned'' to the location of interest, and the
remaining atoms allowed to relax so that the molecule adopted its lowest
energy orientation and internal geometry.  The row of sampled points in
the plane closest to the MOF were found to have energies for all
molecules of more than 2.4~eV greater than the binding site, making it
unphysical for the molecule to sample these regions even under high
temperature and pressure situations.

By definition, at the binding sites the plotted values coincide with the
values in Table~\ref{tab:nmr}. Around the primary binding site in the
energy map we have also indicated the 25.6~meV isoline ($k_BT$ at room
temperature), which gives an estimate for the extent of spatial
fluctuations of the molecules at room temperature. Note that we have
also transferred this ``fluctuation region'' to the NMR maps, where they
can now be used to estimate the fluctuations of the shielding change
that can be expected at room temperature.  However, it turns out that in
all three cases these fluctuations are small.

It is also interesting to see that even in the middle of the MOF all
guest molecules show a shielding significantly different from their gas
phase.  Although at that point the molecules are far away from the
inside wall of the MOF, they are effectively surrounded by twelve
benzene-like linkers (see the setup in Fig.~\ref{fig:loadings}), the
$\pi$ clouds of which influence the electronic structure of the adsorbed 
molecules.  In the case of H$_2$ we investigated this behavior further
with simple model calculations using \textsc{Gaussian} with a
6-311++G(2d,2p) basis set: ``Approximating'' the MOF with 12 benzene
rings and putting the H$_2$ in the center we find a difference in
shielding to the gas phase of 0.91~ppm, whereas the full calculation
with the periodic MOF structure using \textsc{QauntumEspresso} gives
0.92~ppm.

For \Htwo\ and \HtwoO, Figs.~\ref{fig:h2} and \ref{fig:h2o} suggest
qualitatively similar behavior of the shielding for displacements near the
binding site.  However, as indicated in Table~\ref{tab:nmr} and
Figs.~\ref{fig:h2} and \ref{fig:h2o}, while \HtwoO\ binds closer to the MOF
(compare indicated positions of the binding sites in the maps) and deshields
relative to the gas phase, \Htwo\ binds slightly further away, although
when forced closer to the metal ion site it also deshields.

Note that there is not a monotonic increase for either the energy or
shielding when moving the small molecule from the center towards the
MOF. The true function of shielding and energy throughout the cavity of
the MOF is quite complicated, but we provide here a simple model to
approximate both. To this end, we mimic the energy in the plane by a
simple two-dimensional function (of $r$ and $\phi$ in polar coordinates)
inspired by a Lennard-Jones potential with angular dependence, i.\,e.
\begin{equation}
\label{eq:e_model}
\Delta E = \bigg[\frac{a}{(r_0-r)^x} - \frac{a}{(r_0-r)^{x/2}} \bigg] 
           \Big[1 - b \sin^2(f\phi)\Big] + E_0\;.
\end{equation} 
Here, $r$ is the distance from the center of the MOF and $\phi$ is the
polar angle. The zero angle is half-way between two Mg atoms.  It turns
out that a very similar model can also accommodate the spatial change in
NMR shielding fairly well with only a minor modification, although there
is no intuitive reason to believe the shielding should behave radially
in a Lennard-Jones manner.  For the shielding model, we reuse the energy
model with the addition of a radial dependence in the angular piece of
the function,
\begin{equation}
\label{eq:nmr_model}
\Delta \sigma = \bigg[\frac{a}{(r_0-r)^x} - \frac{a}{(r_0-r)^{x/2}} \bigg] 
                \Big[1 - rb \sin^2(f\phi)\Big] + \sigma_0\;.
\end{equation}
These models were tested against the positional data used for the
interpolation shown in Figs.~\ref{fig:h2}--\ref{fig:h2o}, but only the
originally sampled points were used in the parameter fitting.  The
parameters found for all three adsorbed molecules using both models are
shown in Table~\ref{tab:model}, along with the resulting mean absolute
error (MAE) in either units of eV or ppm, as appropriate.

\begin{table}
\caption{\label{tab:model} Parameters for the models described in
Eqs.~(\ref{eq:e_model}) and (\ref{eq:nmr_model}), along with the
resulting mean absolute error (MAE), in either units of eV or ppm as
appropriate, when compared to the original data.}
\begin{tabular*}{\columnwidth}{@{\extracolsep{\fill}}llcccccc@{}r@{}}\hline\hline
               &        & $a$   & $f$   & $x$   & $r_0$ & $b$   & $E_0, \sigma_0$ & MAE \\\hline
$\Delta E$     & \Htwo  & 0.64  & 2.79  & 3.43  & 6.83  & 0.67  & 0.15  & 0.02\\
               & \COtwo & 1.85  & 4.84  & 2.70  & 6.50  & 1.07  & 0.48  & 0.19\\
               & \HtwoO & 5.59  & 3.12  & 10.1  & 6.74  & 2.18  & 0.86  & 0.15\\
$\Delta \sigma$& \Htwo  & --1.76 & --2.02 & 4.24 & 7.27  & 0.93 & 0.98  & 0.19\\
               & \COtwo & 10.1   & 0.61  & 17.8 & 6.75  & --1.84 & 1.01 & 0.21\\
               & \HtwoO & --0.19 & --1.91 & 5.86 & 7.03  & 16.3 & 1.00  & 0.17\\\hline\hline
\end{tabular*}
\end{table}

In both models, there appear to be six free parameters, but some further
simplifications can be made.  First, there is an overall shift parameter
denoted by $E_0$ and $\sigma_0$ in Eq.~(\ref{eq:e_model}) and
Eq.~(\ref{eq:nmr_model}), respectively.  These values can be used in the
fit minimization, but essentially turn out to be the value of the energy
change $\Delta E$ or change in shielding $\Delta \sigma$ at the $r=0$
point in the middle of the MOF.  Hence, these values can be fixed
accordingly in order to reduce the number of tuneable parameters in the
model.  Second, the $r_0$ parameter in all cases is within 1~\AA~of the
distance from the middle of the MOF to the metal ion at the binding site
and so could also be fixed, leaving only four overall adjustable
parameters.  We report in Table~\ref{tab:model} these two parameters in
addition to the remaining four so that the reader can get a sense of the
magnitude and variance of these potentially ``fixable'' pieces of the
model and more easily understand their roles in Eqs.~(\ref{eq:e_model})
and (\ref{eq:nmr_model}).

We also find that the $\Delta \sigma$ model parameters for \COtwo\
reinforce what is shown in Fig.~\ref{fig:co2}, i.e.\ that the shielding
for \COtwo\ adsorbed in MOF-74-Mg behaves qualitatively differently from
\Htwo\ and \HtwoO.  There is also a difference in the local adsorption
geometry.  Whereas both the \Htwo\ and \HtwoO\ molecules adsorb
``flat-on,'' that is, almost parallel to the metal-oxygen plane which
makes up the primary binding site, \COtwo\ adsorbs tilted upwards as
explained in Sec.~\ref{sec:rot}, pointing towards the Mg ion and
interacting to a greater degree with the metal ion itself than with the
oxygen plane.  As pointed out in Ref.~\onlinecite{Valenzano10}, the
\COtwo\ does interact with the oxygens at the binding site so that its
rotation is affected, but it is not ``pulled over'' to the extent that
\Htwo\ and \HtwoO\ are.  Both the qualitative NMR behavior and
adsorption geometry reflect a substantially different MOF-adsorbate
interaction for the \COtwo\ case from the \Htwo\ and \HtwoO\ cases.

\section{Conclusions}

In summary, we have performed \emph{ab initio} DFT simulations of the
energy and NMR chemical shielding of \Htwo, \COtwo, and \HtwoO\ in the
MOF-74-Mg structure.  Our calculations show that there is a 
relationship between loading in MOF-74-Mg and
the change in NMR shielding. While the loading dependence of the
shielding is small, it still is within the measurable experimental
range. We thus argue that combining the NMR signal strength with the
peak positions can yield an accurate tool for determining the loading of
MOFs.  We have further shown how the energy and shielding behave as the
molecules rotate or leave the binding site by providing detailed
energies and NMR shieldings throughout the cavity of the MOF. From our
calculated data, we were able to approximate the energies and shieldings
with two simple functions.  Although our study only investigated one
particular MOF, we believe that the same approach is suitable for other
MOFs and similar studies on e.g.\ MOF-5 and Zn(bdc)(ted)$_{0.5}$ are
already in progress.

\section*{Acknowledgements}

This work was supported in full by the Department of Energy Grant,
Office of Basic Energy Sciences, Materials Sciences and Engineering
Division, Grant No. DE-FG02-08ER46491.



\begin{thebibliography}{99}
\bibitem{foot1} 
     Although the NMR parameters are being calculated for individual
     molecules, our calculations have shown that the so--called $G=0$
     contribution to the magnetic susceptibility related to the macroscopic
     shape of the sample does have a significant effect on the NMR
     results and needs to be included.
\bibitem{foot2} Note that there is a difference of sign convention between
    this work and the work in Ref.~\onlinecite{Thonhauser09}.
\end{thebibliography}
\end{document}